\documentclass[11pt,draftcls,onecolumn]{IEEEtran}
\usepackage{amsfonts}
\usepackage{amssymb}
\usepackage{mathrsfs}
\usepackage{amsmath}
\usepackage{cite}
\usepackage{mathrsfs}
\usepackage[ruled,vlined]{algorithm2e}
\usepackage{tikz}
\usetikzlibrary{arrows,automata}
\usepackage[latin1]{inputenc}
\usepackage{verbatim}
\makeatletter

\newcommand{\Rmnum}[1]{\expandafter\@slowromancap\romannumeral #1@}
\makeatother

\newtheorem{thm}{Theorem}
\newtheorem{lemma}[thm]{Lemma}
\newtheorem{eg}{Example}
\newtheorem{prop}{Proposition}

\newcommand{\w}{{\omega}}
\newcommand{\vb}{\vec{b}}

\newcommand{\vf}{\vec{f}}
\newcommand{\vk}{\vec{k}}

\newcommand{\vm}{\vec{m}}
\newcommand{\Fq}{\mathbb{F}_q}

\newcommand{\mC}{\mathcal{C}}
\newcommand{\mL}{\mathcal{L}}

\newcommand{\bzero}{{\bf 0}}
\newcommand{\wtE}{\widetilde{E}}
\newcommand{\Rank}{{\mathrm{Rank}}}
\newcommand{\tail}{{\mathrm{tail}}}

\newcommand{\In}{{\mathrm{In}}}
\newcommand{\cut}{{\mathrm{cut}}}
\newcommand{\CUT}{{\mathrm{CUT}}}
\newcommand{\mincut}{{\mathrm{mincut}}}
\newcommand{\minord}{{\mathrm{minord}}}
\newcommand{\MinCut}{{\mathrm{MinCut}}}
\newcommand{\mcsim}{\stackrel{\rm{mcut}}{\sim}}

\begin{document}

\title{{\LARGE Small Field Size for Secure Network Coding}
\thanks{This research is supported by the National Key Basic Research Program of China (973 Program Grant No. 2013CB834204), the National Natural Science
Foundation of China (Nos. 61301137, 61171082).}
}
\author{Xuan~Guang~\IEEEmembership{Member,~IEEE},
        Jiyong~Lu~\IEEEmembership{Student Member,~IEEE},
        and~Fang-Wei Fu
\thanks{X. Guang is with the School of Mathematical Sciences and LPMC, J. Lu and F.-W. Fu are with the Chern Institute of
Mathematics, Nankai University, Tianjin 300071, China (e-mail:
xguang@nankai.edu.cn, lujiyong@mail.nankai.edu.cn, fwfu@nankai.edu.cn).}}

\markboth{IEEE COMMUNICATIONS LETTERS}%
{}
{}
%


\maketitle

\begin{abstract}
In network coding, information transmission often encounters wiretapping attacks. Secure network coding is introduced to prevent information from being leaked to adversaries. For secure linear network codes (SLNCs), the required field size is a very important index, because it largely determines the computational and space complexities of a SLNC, and it is also very important for the process of secure network coding from theoretical research to practical application. In this letter, we further discuss the required field size of SLNCs, and obtain a new lower bound. This bound shows that the field size of SLNCs can be reduced further, and much smaller than the known results for almost all cases.
\end{abstract}

\IEEEpeerreviewmaketitle

\section{Introduction}

\IEEEPARstart{I}{n} the paradigm of network coding, when wiretapping attacks occur, that is, an eavesdropper has capability of wiretapping on an unknown channel-set in networks, secure network coding is introduced to prevent information from being leaked to adversaries. This was first proposed by Cai and Yeung in \cite{secure-conference}. In their recent paper \cite{Cai-Yeung-SNC-IT}, they proposed the model of a communication system on a wiretap network (CSWN) and a construction of secure linear network codes (SLNCs) to guarantee that the eavesdropper can obtain no information about the source message and meanwhile all sink nodes as legal users can decode the source message with zero error. Particularly, if the eavesdropper can obtain nothing about the source message by accessing any $r$ channels, we say that this SLNC achieves the security-level $r$. Subsequently, Rouayheb \textit{et al.} \cite{Rouayheb-IT} showed that this model can be regarded as a network generalization of the wiretap channel \Rmnum{2} and presented another construction of SLNCs by applying secure codes for wiretap channel \Rmnum{2}. Motivated by Rouayheb \textit{et al.}s' formulation \cite{Rouayheb-IT}, Silva and Kschischang \cite{Silva-UniversalSNC} studied the universal secure network coding via rank-metric codes, that is, the design of linear network codes for message transmission and the design of security against an eavesdropper can be completely separated from each other. For any construction of SLNCs, the required alphabet size or the size of base finite field, is very important, because it largely determines the complexities of constructions including space and computational complexities, and further efficiency of network transmission. This index is also very important for the process of secure network coding from theoretical research to practical applications. However, the existing results show that these constructions require very large field size, which leads to inefficiency in general. In \cite{Feldman}, Feldman \textit{et al.} derived tradeoffs between security, code alphabet, and information rate of SLNCs, which indicated that if we give up a part of overall capacity, we may use a field of smaller size. This tradeoff can be also obtained from \cite{Cai-Yeung-SNC-IT} as mentioned in their paper.

Motivated by the importance of field size, in this letter, we further explore the required field size for SLNCs, and give a new lower bound. This bound shows that the field size can be reduced considerably further by applying network topologies without giving up any part of capacity. This is benefit to the implementation of SLNCs in possible applications. To be specific, by observing the constructions of SLNCs in \cite{Cai-Yeung-SNC-IT, Rouayheb-IT, Silva-UniversalSNC}, we introduce a useful equivalence relation in networks, and the number of equivalence classes induced by this equivalence relation is sufficient for the construction of SLNCs. Finally, an example is illustrated to compare our result with the previous.

\section{The Field Size of SLNCs}
\subsection{Related Works}
First, we state the construction proposed by Cai and Yeung \cite{Cai-Yeung-SNC-IT} for designing an $\w$-rate and $r$-security-level SLNC on a single source multicast network $G$ with unit capacity channels. In addition, Rouayheb \textit{et al.} \cite{Rouayheb-IT} proposed another construction and indicated that their construction is actually equivalent to Cai and Yeung's.
\newline \noindent \textbf{Construction:}
\newline \noindent \textbf{Step 1:}\quad For the information rate $\w$ and security-level $r$ with $n\triangleq \w+r\leq C_{\min}=\min\limits_{\text{all sinks }t}C_t$ with $C_t$ being the minimum cut capacity between the single source $s$ and the sink $t$, construct an $n$-dimensional $\Fq$-valued linear network code (LNC) $\mC_n$ of global encoding kernels $\vf_e$, $e\in E$, (e.g. Jaggi \emph{et al.}'s algorithm \cite{co-construction});
\newline \noindent \textbf{Step 2:}\quad Choose $n$ linearly independent column vectors $\vb_1,\ \vb_2,\ \cdots,\ \vb_n\in \Fq^n$ satisfying the \textit{secure condition}:
  \begin{equation}\label{secure_condition}
  \langle \vb_i:\ 1\leq i \leq \w \rangle \cap \langle \vf_e:\ e\in A \rangle=\{\bzero\}\footnote{Let $L$ represent a collection of vectors in some linear space, and then we use $\langle L \rangle$ to denote the subspace spanned by vectors in $L$ for convenience. In addition, we always use $\bzero$ to denote all zero column vectors throughout this letter, whose dimensions will always be clear from the context.}
  \end{equation}
  for all channel-subsets $A\in E$ of cardinalities no larger than $r$. Then define an $n\times n$ invertible matrix
  $Q=\big[ \vb_1 \ \vb_2 \ \cdots \ \vb_n \big]$.
\newline \noindent \textbf{Step 3:}\quad The source message $M$ is randomly chosen from $\Fq^\w$ and the independent random key $K$ is distributed uniformly on $\Fq^r$, both of which are generated only at the source. Let an $\w$-row vector $\vm\in \Fq^{\w}$ and an $r$-row vector $\vk\in \Fq^{r}$ be the outcomes of $M$ and $K$, respectively. Thus $\vec{x}=(\vm,\ \vk )$ is the input of the network, and then send the pre-encoded input $\vec{x}\cdot Q^{-1}$ through the network by using the LNC $\mC_n$.

This construction designs an $\w$-rate and $r$-security-level SLNC. Actually, $\{ Q^{-1}\cdot\vf_e:\ e\in E \}$ constitutes a global description of this SLNC.
Evidently, the secure condition (\ref{secure_condition}) must be qualified for all $A\subseteq E$ of cardinalities no larger than $r$ if it is qualified for those
$A\subseteq E$ with $|A|=r$. Hence, we just need to consider the number of channel-sets $A$ of cardinality $r$, equal to ${|E| \choose r}$. Hence, we give Cai and Yeung's conclusion \cite{Cai-Yeung-SNC-IT} on the required field size below.
\begin{thm}
A $\Fq$-valued SLNC with rate $\w$ and security-level $r$ can be constructed provided that $|\Fq|=q\geq {|E| \choose r}$.\footnote{Actually, as shown in \cite{co-construction}, for the LIF algorithm, a field of size larger than or equal to ${|E|\choose r}$ is sufficient for guaranteeing the existence of SLNCs.}
\end{thm}
\vspace{-1em}
\subsection{Our Contributions}
By further observing the secure condition (\ref{secure_condition}), we found that it is sufficient to consider those channel-sets $A\subseteq E$ satisfying $|A|=\Rank(F_A)=r$ with $F_A=\big[ \vf_e: e\in A \big]$, since for any channel-set $A'\subseteq E$ with $|A'|=r$ but $\Rank(F_{A'})<r$, we can always find a channel-set $A\subseteq E$ satisfying $|A|=\Rank(F_A)=r$ such that $\mL_{A'}\subset \mL_{A}$, where $\mL_B=\langle  \vf_e: e\in B  \rangle$ for arbitrary channel-set $B$. Hence, we define a collection $\wtE_r$ consisting of all such $A$ as:
$$\wtE_r \triangleq \{ A\subseteq E:\ |A|=\Rank(F_A)=r \}.$$
The above observation implies that the required field size can be reduced to $|\wtE_r|$. However, notice that $\wtE_r$ depends on the underlying LNC and so it is hard to handle. Naturally, we hope to find another lower bound on the field size just depending on network topologies. Thus, we define a new collection of channel-sets as:
\begin{align}\label{def_col-cut}
\wtE_r^{\cut} \triangleq \{ A\subseteq E:\ |A|=\mincut(s, A)=r \},
\end{align}
where $\mincut(s, A)$ represents the minimum cut capacity between $s$ and $A$, and clearly $\wtE_r^{\cut}$ just depends on the network topology. We first interpret some concepts just mentioned or to be used in the following. Again let $G=(V, E)$ be a finite acyclic directed network with unit capacity channels and the single source $s$, and let $A\subseteq E$ be a channel-set in $G$. At first, we define a cut between $s$ and $A$ in $G$. In the network $G$, install a new node $t_A$, and for every edge $e\in A$, add a new edge $e'$ connected from $\tail(e)$ to the new node $t_A$ and meanwhile delete the edge $e$ from $G$. A cut between $s$ and $t_A$ is regarded as a cut between $s$ and $A$ in $G$, but it is necessary to mention that if a cut separating $s$ and $t_A$ contains some edges in $\In(t_A)$, then they should be replaced by the corresponding edges in $A$. Further, the minimum cut capacity between $s$ and $A$ is defined as the minimum cut capacity between $s$ and $t_A$, and the cuts separating $s$ and $A$ achieving this minimum cut capacity are called the minimum cuts. Now, we can obtain the following proposition easily, which implies that the size of $\wtE_r^{\cut}$ can be regarded as a new lower bound.
\begin{prop}
$\wtE_r \subseteq \wtE_r^{\cut}$, \ \ and \ \ $|\wtE_r|\leq |\wtE_r^{\cut}| \leq { |E| \choose r}$.
\end{prop}

Thus, we can obtain our first conclusion.
\begin{thm}
A $\Fq$-valued SLNC with rate $\w$ and security-level $r$ can be constructed provided that $|\Fq|=q \geq |\wtE_r^{\cut}|$.
\end{thm}

As we mentioned above, what we are concerned are those vector spaces spanned by global encoding kernels $\vf_e$, $e\in A$, i.e., $\mL_A=\langle \vf_e: e\in A \rangle$ for all $A \in \wtE_r$, or further all $A \in \wtE_r^{\cut}$. Furthermore, notice a fact that in linear network coding, for any channel-set $A$, all global encoding kernels of channels in $A$ are linear combinations of those global encoding kernels of channels in any cut $\CUT$ separating $s$ and $A$. This subsequently means that $\mL_A$ must be a subspace of $\mL_{\CUT}$. Particularly, if $A\in \wtE_r^{\cut}$ and $\CUT$ is a minimum cut between $s$ and $A$, then $\mL_A\subseteq\mL_{\CUT}$. Thus, the number of different vector spaces amongst all vector spaces $\mL_A$ for all $A \in \wtE_r$ or $A \in \wtE_r^{\cut}$ is enough for the above Construction. Motivated by this observation, it is necessary to continue discussing the required finite field in order to reduce its size.

First, notice the following fact that $\mL_A=\mL_{A'}$ for any two channel-sets $A, A'\in \wtE_r$ provided that $A$ and $A'$ have a common minimum cut. In addition, similar to what we indicated before, we still want to find a lower bound on the field size just depending on network topologies. Hence the following discussion is given. At first we define a relation ``$\mcsim$'' between arbitrary two channel-sets $A$ and $A'$ in $\wtE_r^{\cut}$:
\begin{align}\label{def_mcut-relation}
A \mcsim A'
\end{align}
if and only if $A$ and $A'$ have a common minimum cut between the source node $s$ and them. The theorem below shows the relation ``$\mcsim$'' in $\wtE_r^{\cut}$ being an equivalence relation.
\begin{thm}\label{thm_equiv-relation}
The relation ``$\mcsim$'' is an equivalence relation. Equivalently, the following three properties are qualified for all channel-sets $A, A', A'' \in \wtE_r^{\cut}$:
\begin{enumerate}
  \item {\bf(Reflexivity)} $A\mcsim A$;
  \item {\bf(Symmetry)} if $A\mcsim A'$ then $A' \mcsim A$;
  \item {\bf(Transitivity)} if $A\mcsim A'$ and $A'\mcsim A''$, $A\mcsim A''$.
\end{enumerate}
\end{thm}

The reflexivity and symmetry of the relation ``$\mcsim$'' are obvious. To show transitivity, we need two lemmas below.
\begin{lemma}\label{app_lem-1}
Let $G=(V,E)$ be a finite acyclic directed network with unit capacity channels and the single source $s$. Let $t$ be a non-source node and the minimum cut capacity between $s$ and $t$ be $r$. Then arbitrary $r$ edge-disjoint paths from $s$ to $t$ pass through all minimum cuts between $s$ and $t$, and $r$ distinct edges in each minimum cut are on $r$ distinct paths respectively.
\end{lemma}
\begin{IEEEproof}
Assume the contrary that there exists a minimum cut $\CUT$ between $s$ and $t$ such that there is an edge $e\in \CUT$ on none of $r$ paths. Note that $\CUT$ is minimum, and so $|\CUT|=\mincut(s,t)=r$, which shows that all edges in $\CUT$ are on $r-1$ edge-disjoint paths at most. Thus, there must exist one path passing through no edges in $\CUT$. This further implies that after deleting all edges in $\CUT$ from the network $G$, there still exists an path from $s$ to $t$, which conflicts with the assumption that $\CUT$ is a cut between $s$ and $t$.
\end{IEEEproof}

For any two edges $e_i$ and $e_j$ in a directed acyclic network, if a path from $e_i$ to $e_j$ can be found, then we say that $e_i$ is previous to $e_j$, denoted by $e_i\prec e_j$. Particularly, we set $e\prec e$ for every edge $e$. This is a natural and conventional partial order in directed acyclic networks.

\begin{lemma}\label{app_lem-2}
Let $G=(V,E)$ be a finite acyclic directed network with unit capacity channels and the single source $s$, and let $t$ be a non-source node of the minimum cut capacity $r$ from $s$. Further let $P_1, P_2, \cdots, P_r$ be $r$ edge-disjoint paths from $s$ to $t$, and $\CUT_1=\{ e_{1,i}: 1\leq i \leq r \}$ and $\CUT_2=\{ e_{2,i}: 1\leq i \leq r \}$ be two minimum cuts between $s$ and $t$ with $e_{j,i}$ on the path $P_i$ for $1 \leq i \leq r$, $j=1, 2$. Define an edge-set $\CUT=\{ \minord(e_{1,i}, e_{2,i}): 1\leq i \leq r \}$, where
\begin{align*}
\minord(e_{1,i}, e_{2,i})=
\begin{cases} e_{1,i}, & e_{1,i} \prec e_{2,i},\\
              e_{2,i}, & \text{otherwise.}
\end{cases}
\end{align*}
Then $\CUT$ is still a minimum cut between $s$ and $t$.
\end{lemma}
\begin{IEEEproof}
First, since $|\CUT|=r=\mincut(s,t)$, $\CUT$ must be a minimum cut between $s$ and $t$ provided that it is a cut between them. Hence, we will just show that $\CUT$ is a cut between $s$ and $t$. Two cases below are discussed.

\noindent\textit{\textbf{ Case 1. }} If $\minord(e_{1,i}, e_{2,i})=e_{1,i}$ (resp. $e_{2,i}$), i.e., $e_{1,i}\prec e_{2,i}$ (resp. $e_{2,i}\prec e_{1,i}$) for all $1\leq i \leq r$, then the result of the theorem is trivial.

\noindent\textit{\textbf{ Case 2. }} Otherwise, assume the contrary that $\CUT$ is no longer a cut between $s$ and $t$. Then there must exist a path $P$ from $s$ to $t$ which doesn't pass through $\CUT$. But this path $P$ has to pass through the minimum cut $\CUT_1$ from Lemma \ref{app_lem-1}, and let $e_{1,i_1}$ in $\CUT_1$ be the channel passed through by $P$. Then we can claim that $e_{1,i_1} \succ e_{2,i_1}$. Conversely, if $e_{1,i_1} \prec e_{2,i_1}$ then $\minord(e_{1,i_1}, e_{2,i_1})=e_{1,i_1}\in \CUT \cap P$, which leads to a contradiction. Further, we replace the part from $e_{1,i_1}$ to $t$ in $P$ by the part from $e_{1,i_1}$ to $t$ in $P_{i_1}$, and keep the remaining part in $P$, i.e., the part from $s$ to $e_{1,i_1}$, unchanged. Then we derive a new path denoted by $P^{(1)}$. It is easy to check that $P^{(1)}$ doesn't pass through $\CUT$ either, since none of the parts from $s$ to $e_{1,i_1}$ in $P$ and from $e_{1,i_1}$ to $t$ in $P_{i_1}$ pass through $\CUT$.

Now, we claim that the sub-path from $s$ to $e_{1,i_1}$ of $P^{(1)}$ must pass through the minimum cut $\CUT_2$. Since the part from $e_{1,i_1}$ to $t$ in $P^{(1)}$ (the same as the corresponding part in $P_{i_1}$) cannot pass through $\CUT_2$ from $e_{1,i_1}\succ e_{2,i_1}$, the path $P^{(1)}$ would not pass through $\CUT_2$ provided that its sub-path from $s$ to $e_{1,i_1}$ doesn't pass through $\CUT_2$. This contradicts to Lemma \ref{app_lem-1}. Thus, assume that the sub-path from $s$ to $e_{1,i_1}$ of $P^{(1)}$ passes through $e_{2,i_2}$ in $\CUT_2$, and similarly, $e_{2,i_2}\succ e_{1,i_2}$, otherwise $\minord(e_{1,i_2}, e_{2,i_2})=e_{2,i_2}\in P^{(1)}\cap \CUT$ yielding a contradiction. We further replace the part from $e_{2,i_2}$ to $t$ in $P^{(1)}$ by the part from $e_{2,i_2}$ to $t$ in $P_{i_2}$, and keep the part from $s$ to $e_{2,i_2}$ in $P^{(1)}$ unchanged. Then we again construct a new path $P^{(2)}$ from $s$ to $t$, which doesn't pass through $\CUT$. Subsequently, note that the length of the sub-path from $s$ to $e_{2,i_2}$ of the path $P^{(2)}$ is strictly smaller than the length of the sub-path from $s$ to $e_{1,i_1}$ of the path $P^{(1)}$ because the latter covers the former and the network is acyclic.

By the same analysis as above, the path from $s$ to $e_{2,i_2}$ in the path $P^{(2)}$ must pass through some channel $e_{1,i_3}$ in $\CUT_1$ and $e_{1,i_3}\succ e_{2,i_3}$. Replace the part from $e_{1,i_3}$ to $t$ in $P^{(2)}$ by the part from $e_{1,i_3}$ to $t$ in $P_{i_3}$ and keep the part from $s$ to $e_{1,i_3}$ in  $P^{(2)}$ unchanged, which constitutes a new path $P^{(3)}$ from $s$ to $t$ which doesn't pass through $\CUT$. Moreover, the length of the sub-path from $s$ to $e_{1,i_3}$ in the path $P^{(3)}$ is strictly smaller than the length of the sub-path from $s$ to $e_{2,i_2}$ in the path $P^{(2)}$. So far and so forth, because the length of the sub-path from $s$ to $e_{1,i_1}$ in the path $P$ is finite, this process will stop at some step. This implies that finally we find a path from $s$ to $t$ doesn't pass through either $\CUT_1$ or $\CUT_2$, which is a contradiction. Therefore, our hypothesis is not true, i.e., $\CUT$ is also a cut further a minimum cut between $s$ and $t$.
\end{IEEEproof}
\begin{IEEEproof}[Proof of Theorem \ref{thm_equiv-relation}]
Review the collection $\wtE_r^{\cut}$ in (\ref{def_col-cut})
and the relation ``$\mcsim$'' from (\ref{def_mcut-relation}).
We will prove the transitivity of the relation ``$\mcsim$''.

First, for any $A\in \wtE_r^{\cut}$, define $\MinCut(A)$ as the collection of all minimum cuts between $s$ and $A$. Let $\CUT_1$ be a common minimum cut of $A$ and $A'$, i.e., $\CUT_1\in \MinCut(A) \cap \MinCut(A')$, and similarly $\CUT_2$ be a common minimum cut of $A'$ and $A''$, i.e., $\CUT_2\in \MinCut(A') \cap \MinCut(A'')$. Further, by Menger's Theorem, we can find $r$ edge-disjoint paths $P_1, P_2, \cdots, P_r$ from $s$ to $A'$, and no other paths from $s$ to $A'$ exist provided deleting these $r$ paths. Together with Lemma \ref{app_lem-1}, these $r$ paths pass through all minimum cuts in $\MinCut(A')$, and particularly, pass through $\CUT_1$ and $\CUT_2$. Subsequently, let $\CUT_j=\{ e_{j,i}: 1 \leq i \leq r \}$, $j=1,2$, and let $e_{j,i}$ be on the path $P_i$ for all $1 \leq i \leq r$ and $j=1,2$. Next, define a new channel-set $\CUT$:
\begin{equation*}
\CUT=\{\minord(e_{1,i}, e_{2,i}):\ 1\leq i \leq r \}.\footnote{Actually, $\CUT$ is independent with the choice of the $r$ paths from $s$ to $A'$.}
\end{equation*}
By Lemma \ref{app_lem-2}, we know that $\CUT$ is still a minimum cut between $s$ and $A'$, that is, $\CUT\in \MinCut(A')$. In the following, we will prove that $\CUT$ actually is a common minimum cut between $s$ and both $A$ and $A''$. In other words, $\CUT\in \MinCut(A)\cap\MinCut(A'')$.

At first, we show $\CUT\in\MinCut(A)$. Conversely, $\CUT \notin \MinCut(A)$. Then there still exists a path $P_A$ from $s$ to $A$ after deleting all edges in $\CUT$ from $G$. Subsequently, this path $P_A$ must pass through $\CUT_1$ since $\CUT_1$ is a minimum cut between $s$ and $A$. Without loss of generality, assume that $P_A$ passes through the edge $e_{1,1}$ in $\CUT_1$. Then $e_{1,1}\succ e_{2,1}$ because $e_{1,1}\notin \CUT$ and $P_A$ doesn't passes through $\CUT$. Furthermore, since $\CUT_1$ is also a minimum cut separating $s$ and $A'$, there is a path from $e_{1,1}$ to $A'$. Now, we can construct a path $P_{A'}$ from $s$ to $A'$ constituted by the part from $s$ to $e_{1,1}$ in $P_A$ concatenated by the part from $e_{1,1}$ to $A'$ in $P_1$. Notice none of the two parts passing through $\CUT$. This means that $P_{A'}$ doesn't pass through $\CUT$, conflicting with the fact that $\CUT$ is a minimum cut separating $s$ and $A'$. Thus, our hypothesis is not true and it follows $\CUT\in \MinCut(A)$.

Similarly, we can also prove $\CUT\in \MinCut(A'')$. Combining the above, we obtain $A \mcsim A''$. Therefore, we complete the proof of the transitivity.
\end{IEEEproof}

Therefore, the relation ``$\mcsim$'' can give a partition of $\wtE_r^{\cut}$ because of its equivalence property, and the number of all equivalence classes induced by ``$\mcsim$'' forms a new lower bound on the size of the required finite field for Construction, which is our main conclusion in this letter.
\begin{thm}\label{thm_main}
A $\Fq$-valued SLNC with rate $\w$ and security-level $r$ can be constructed provided that the field size $|\Fq|$ is larger than or equal to the number of equivalence classes of $\wtE_r^{\cut}$ induced by the equivalence relation ``$\mcsim$''.
\end{thm}

In addition, Rouayheb \textit{et al.} \cite{Rouayheb-IT} modified the LIF algorithm, proposed by Jaggi \textit{et al.} \cite{co-construction} for constructing LNCs, to obtain a SLIF algorithm for constructing SLNCs. Although their construction is equivalent to Cai and Yeung's in \cite{Cai-Yeung-SNC-IT}, their bound on the required field size of a SLNC is smaller. To be specific, for their SLIF algorithm, a SLNC with rate $\w$ and security-level $r$ can be constructed over the finite field $\Fq$ of size $q\geq {|E|-1 \choose r-1}+|T|$, and further, combining the algorithm in \cite{Langberg-computational-perspective}, which uses the concept of encoding edges, with their SLIF algorithm, the corresponding field size has to satisfy $|\Fq|=q\geq {2C_{\min}^3|T|^2-1 \choose r-1}+|T|$, which is independent to the number of channels in $G$. Actually, easily see that for almost all cases, our bound in Theorem \ref{thm_main}, the number of equivalence classes, is much smaller than Rouayheb \textit{et al.}s' two bounds. Particularly, we can also apply the algorithm in \cite{Langberg-computational-perspective} to construct an auxiliary network $\hat{G}$ from $G$ firstly, and then apply our analysis on $\hat{G}$. Moreover, it is necessary to notice that the above latter bound by Rouayheb \textit{et al.} \cite{Rouayheb-IT} is worse than the former one in many cases. In addition, motivated by Rouayheb \textit{et al.}s' formulation, universal secure network coding based on packet transmission is discussed in \cite{Silva-UniversalSNC}. Silva and Kschischang show that this universal property can be qualified only if the packet length $m$ must be larger than or equal to the capacity $C_{\min}$, and assume that the eavesdropper wiretaps $r$ packets transmitted on $r$ channels for every wiretap. Actually, their crucial idea is to consider the secure linear code at the source node over an extension field $\mathbb{F}_{q^m}$, and thus the condition (\ref{secure_condition}) can be satisfied for all possible $r\times n$ matrices $F_A^\top$. Hence, it is conceivable that the required field size is larger than ours.
Next, we will give an example to compare our bounds with the existing results.
\begin{eg}\label{eg_com_netw_1}
We take the combination network $G_1$ (see \cite[p.26]{Zhang-book}) with parameters $N=8$ and $k=6$ as an example. To be specific, $G_1$ has a single source $s$ and $N=8$ internal nodes, each of which is connected from $s$ by one and only one channel. Arbitrary $k=6$ internal nodes are connective with one and only one sink node, and so the number of sink nodes is $|T|={N \choose k}={8 \choose 6}=28$, the number of internal nodes is $|J|=8$, the number of channels is $|E|=N+|T|\cdot k=8+28\times6=176$, and evidently the minimum cut capacity $C_{\min}$ between $s$ and every sink is $6$. The Figure \ref{fig_cn} is an illustration of a combination network with $N=3$ and $k=2$.
\begin{figure}[!htb]
\centering
\begin{tikzpicture}
[->,>=stealth',shorten >=1pt,auto,node distance=2cm,
                    thick]
  \tikzstyle{every state}=[fill=none,draw=black,text=black,minimum size=6mm]
  \node[state]         (s)                 {};
  \node[state]         (i_2)[below of=s]   {};
  \node[state]         (i_1)[left of=i_2]  {};
  \node[state]         (i_3)[right of=i_2] {};
  \node[state]         (t_1)[below of=i_1] {};
  \node[state]         (t_2)[below of=i_2] {};
  \node[state]         (t_3)[below of=i_3] {};
\path
(s) edge           node {} (i_1)
    edge           node {} (i_2)
    edge           node {} (i_3)
(i_1) edge node{}(t_1)
      edge node{}(t_2)
(i_2) edge node{}(t_1)
      edge node{}(t_3)
(i_3) edge node{}(t_2)
      edge node{}(t_3);
\end{tikzpicture}
\caption{Combination Network with $N=3,k=2$.}
\label{fig_cn}
\end{figure}
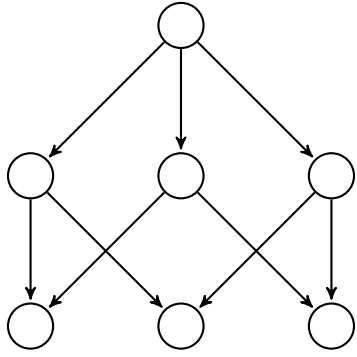

Let the information rate and the security-level be $\w=3$ and $r=3$, respectively. After a simple calculation, we have
${|E| \choose r}={176 \choose 3}=893200.$
Subsequently, we analyze the cardinality of the collection
$\wtE_r^{\cut}=\wtE_3^{\cut}=\{ A\subseteq E:\ |A|=\mincut(s, A)=3 \}.$
We divide all channels in $E$ into two layers: upper layer and lower layer. The upper layer contains all channels between $s$ and internal nodes, and thus there are total $8$ channels in this layer. The lower layer consists of all channels between internal nodes and sink nodes, and total $|T|\cdot k=28 \times 6=168$ channels in this layer. Next we will count the number of channel-sets $A$ in $\wtE_3^{\cut}$.

\begin{itemize}
  \item {\bf \textit{Case1.}}\ All three channels of $A$ are from the upper layer. Then the number of such $A$ in $\wtE_3^{\cut}$ is ${8 \choose 3}=56$.
  \item {\bf \textit{Case2.}}\ All three channels of $A$ are from the lower layer. Notice that if three channels in the lower layer achieve capacity $3$, then they have to come from different internal nodes. Together with the fact that the number of outgoing channels of every internal node is ${N-1 \choose k-1}={7 \choose 5}=21$, the number of such $A$ in $\wtE_3^{\cut}$ is ${8 \choose 3}{21 \choose 1}{21 \choose 1}{21 \choose 1}=518616$.
  \item {\bf \textit{Case3.}}\ Two channels of $A$ are from the upper layer and the other one is from the lower layer. The number of such $A$ in $\wtE_3^{\cut}$ is ${8 \choose 2}{6 \choose 1}{21 \choose 1}=3528$.
  \item {\bf \textit{Case4.}}\ Two channels of $A$ are from the lower layer and the other one is from the upper layer. The number of such $A$ in $\wtE_3^{\cut}$ is ${8 \choose 1}{7 \choose 2}{21 \choose 1}{21 \choose 1}=74088$.
\end{itemize}
Combining the above four cases, one obtains $|\wtE_3^{\cut}|=596288$, smaller than ${|E| \choose r}=893200$.

Next, we focus on the number of equivalence classes in $\wtE_3^{\cut}$ under the relation ``$\mcsim$''. It is easy to deduce that any channel-set $A$ in Cases 2, 3 and 4 must have a minimum cut in Case 1. Thus, the number of the equivalence classes is ${8 \choose 3}=56$. This indicates that the required field size $56$ is enough, which is much smaller than ${|E| \choose r}=893200$ and $|\wtE_r^{\cut}|=596288$.

Further, for the two bounds on the required field size in \cite{Rouayheb-IT}, we calculate that ${|E|-1 \choose r-1}+|T|=15253$ and ${2 C_{\min}^3|T|^2-1 \choose r-1}+|T|>5\times 10^{10}$. Clearly, our new bound is also much smaller than them. In addition, although the authors in \cite{Silva-UniversalSNC} considered the packet network coding problem, and by contrast, we (also \cite{Cai-Yeung-SNC-IT} and  \cite{Rouayheb-IT}) studied scalar network coding problem, we compare our results with theirs. It is known that the field size which can be chosen as the minimum required for multicasting is larger than or equal to the number of sink nodes, i.e., $|T|=28$ here, and further the minimum cut capacity $C_{\min}$ between $s$ and each $t\in T$ is $6$. Thus, the field size as discussed in \cite{Silva-UniversalSNC} for their universal secure scheme is at least $28^6$, also much larger than our result $56$.
\end{eg}


\end{document}